\documentclass[runningheads]{llncs} 

\usepackage{times}
\usepackage{soul}
\usepackage{url}
\usepackage[hidelinks]{hyperref}
\usepackage[utf8]{inputenc}
\usepackage[small]{caption}
\usepackage{graphicx}
\usepackage{amsmath}
\usepackage{booktabs}
\usepackage{algorithm}
\usepackage{algorithmic}
\usepackage{amssymb}
\usepackage{multirow}
 
\DeclareMathOperator*{\mode}{mode}

\begin{document}

\title{A Measurement of Social Capital in an Open Source Software Project} 
\titlerunning{Measuring Social Capital in an OSS Project}
%
\author{ 
Saad Alqithami\inst{1} \and
Musaad Alzahrani\inst{1} \and
Fahad Alghamdi\inst{1,3} \and
Rahmat Budiarto\inst{1} \and
Henry Hexmoor\inst{2} %
}
\authorrunning{S. Alqithami et al.}
%
\institute{Computer Science Department, Albaha University, Saudi Arabia \and
Computer Science Department, Southern Illinois University, Carbondale, IL USA \and
Computer Science Department, George Washington University, Washington DC, USA
\footnotetext{- Corresponding author: \email{salqithami@bu.edu.sa}}
}

\maketitle

\begin{abstract}
The paper provides an understanding of social capital in organizations that are open membership multi-agent systems with an emphasis in our formulation on the dynamic network of social interaction that, in part, elucidate evolving structures and impromptu topologies of networks. This paper, therefore, models an open source project as an organizational network. It provides definitions of social capital for this organizational network and formulation of the mechanism to optimize the social capital for achieving its goal that is optimized productivity. A case study of an open source \texttt{Apache-Hadoop} project is considered and empirically evaluated. An analysis of how social capital can be created within this type of organizations and driven to a measurement for its value is provided. Finally, a verification on whether the social capital of the organizational network is proportional towards optimizing their productivity is considered.
\keywords{Social Capital  \and Open Multi-agent Systems\and Interaction.}
\end{abstract}

\section{Introduction}

An ad-hoc organization of networked agents may form to rally around a specific problem. We explore the effects resultant from networking by addressing one type of network effects called \textit{Social Capital} (SC). 
Social capital in a cross-organizational network can be characterized as collocated or virtual collaboration to produce successful outcomes and successful connections. There are two major perspectives on SC in networks. In the macroscopic perspective, SC for the entire network is considered. In this view individuals do not incrementally add to the system or withdraw units of SC. Instead the foci are on the system principles like norms and conventions that provide resources for overall social welfare. In contrast, the microscopic perspective adopted here explores how individuals can gain access to resources by their positions and connections in the network~\cite{lin1999building}. 
%
Quantities of SC can be used to replace interpersonal trust among agents and that is due to when an organization generates positive values of SC, constituent agents gain benevolence and behave in a trusting manner~\cite{Smith:2009}. 
Other benefits of SC are enhanced group communication, efficient use of intellectual capital, better collective action and easy way to accessing resources~\cite{mendez2009returns}.

There are a few main elements of the SC that have a proportional relationship to one another. %
Topologically speaking, high bonding rates provide more opportunities for interaction and growth of SC. However, network structure by itself is inadequate for the determination of SC. We must examine the contents of interaction and dispositions that create social forces that attract or repel individuals~\cite{hsu}. At the level of a single link, the nature of social flow (i.e., information flow) in the link leads to accumulation of SC. Social flows can be benevolent and positive or negative and lack benevolence. Whereas positive flow leads to network positive gain in SC, negative flow leads to loss of SC. Apart from social flow; dyadic ties may harbor trust or promote distrust. 
Trust supports SC whereas distrust erodes it. If the topic of interactions between the pair is centered on the main problem for an organization, that link positively contributes to SC. Thus, flow, trust, and topic are link attributes that are proportional determinants for SC. 

\emph{Social capital in a link} is the accumulation of positive values of social flow and trust plus abundance of communication over a common topic. 
Since considering a topic of interaction is included in the determination of SC over the link, we note that this formulation of social capital is relativized for links only in an organization. 
SC is generated in the links through dynamics of interaction on the links. Thus, SC for a network linearly scales by summing SC for all links in the network. Increased links are proportional to increase social capital (i.e., network bonding measure). The effects of topology are overlooked in this network perspective but will be considered egocentrically. 
From an egocentric perspective, bridging is said to contribute to social capital~\cite{Smith:2009}. 
Network bonding leads to increased density and closure in the network, which increases resource access~\cite{lin1999building}. 
%
%
%
%
\emph{Social capital for an agent} is the egocentric for an individual that deliberately mirrors the Bonacich Power Index \cite{Bonacich:87}. This coincidence helps us to exploit the topological position of nodes. An agent that is well positioned by having a High Power Index (i.e. high Bonacich centrality value~\cite{Bonacich:87}) will similarly possess high SC.

The previous definitions and views of SC in a restrictive network open the discussion to consider them in organizations. Organizations, in general, are bounded networks with purposeful interaction between their agents. 
Organizations have multiple degrees of institutionalized culture, norms and values that are essential in the development of SC. 
SC receives direct and indirect effects from formal institutions due to formal relations that have been provided by the organization to create interpersonal relations which contribute positively to SC~\cite{Adler:2000}.
SC in an organization represents the resultant outcome from SCs embedded in a social system or through a direct or indirect social relations of an agent, including inherited norms and culture values. Values of SC are built within the social structure to facilitate the agents' actions and interaction~\cite{Adler:2000}.

This paper attempts to measuring SC of an \textit{Open Source Software} (OSS) project and shows that SC value changes proportionally to the goal achievements of the OSS project. 
GitHub is the most popular platform for open source collaboration. On GitHub, developers can join and contribute to projects by submitting issues and contributing code. Submitting issues means sending messages about errors in applications and suggesting ways to fix them. Contributing code involves sending pull requests with the corrections and improvements. A project team is considered as an organizational network, which consists of developers as nodes and each one may have relations with others through a common tasks in modules.


\section{Related Work}
Social capital has been studied by many previous researchers~\cite{Gummer:1998,Nahapiet:1998,Burt:2000-NetSC,
Pitt2016}; 
however, a unified definition of it is a critical issue. 
Bourdieu 
\cite{Bourdieu:1986} refers to SC as the actual or potential collective resources in an institutionalized synergistic network of homogenous agents which in some cases may result into other forms of capitals. 
The point behind SC is to make use of the accumulation of resources embedded in the social structure~\cite{lin2002social}. 
Other authors \cite{ostrom:2003} have defined SC as an attribute of individuals that enhances their abilities to solve collective action problems. Furthermore, Nahapiet and Ghoshal \cite{Nahapiet:1998} described SC through three different dimensions: 
1) \textit{structure dimension} to include the properties of the whole network, 
2) \textit{relational dimension} to present the values of exchanges in agents' connections, and 
3) \textit{cognitive dimension} to support the homogeneity by sharing interpretations and mutual understanding between agents. 

There are two types of social capital in an intra-organizational network: bonding and bridging. 
Both types are generated from agents' interaction (i.e. network homophily
~\cite{mcpherson2001birds}). 
A major difference between those two types is that bonding SC occurs between homogenous agents working on a common goal while bridging involves interaction between heterogeneous agents who are not necessarily working for the same goal~\cite{lin2002social}. Bonding SC increases through closure which contributes positively to the values of relations. Although bridging SC can be considered between agents within an organization, increase of its value can, in some cases, be a resultant of interaction through an inter-organizational network and bridging a cross structural holes~\cite{burt2005brokerage}. 
In short, the literature is a good addition to our view of SC, yet it falls short in differentiating SC from social network analysis that is directly affected by the network topologies. An analyses on the structural dimension (e.g.,  asymmetric emerging distribution of interrelations) of social capital considering the impact of it on the success of open source software projects have been discussed in~\cite{mendez2009returns}.


The rich literatures on social capital has provided the approach with quite valuable components. The deployment of those
parameters into this work in order to efficiently measure and exploit social in an OSS results in several benefits. This
is because traditional studies on social capital consider only the total number of ties an individual or organization has, ignoring the direction of the social flows. In our measurements, however, we signify the sources of social capital by considering reciprocity exchange theory to measure it. Besides, we consider the impact of SC on a real-world case study of an OSS project, which has set it apart from traditional prior techniques.

\section{Quantifying Social Capital}

We consider SC to be a scalar value that can be accumulated as well as consumed either verbatim or used as credit. In a network, SC might be used to trade for help or exchanges with others in the form of delegation of tasks. Bartering with SC can be limited to a pair of agents through an immediate link between them. Alternatively, an agent might enter bartering anonymously with another agent with whom there may not be a directed relationship. 

Our measurements of SC on OSS Project is based on a weighted task-based directed graph inherited from the
general dependency-network graph. An organization (i.e. the OSS Project in our case) is modeled as a directed
graph of agents (contributors) as vertices and their cumulative values of relations between the contributors as edges:
$\{\mathcal{N}, R_{elation}\}$, where $\mathcal{N}$ is the set of agents in an organization that is $\geq 2$, and $R_{elation} \subseteq \mathcal{N} \times \mathcal{N}$ is the set of directed relations between agents. The organization has a common goal that is divided into a set of tasks. Each task will be conducted by a subgroup of ${N} \subseteq \mathcal{N}$. Each agent has a capacity extracted from her public profiles, which include capability, willingness and previous relations.

\subsection{Parameters of SC in an SSO Project}
We propose a measurement of relations from continual interaction and a quantification of an agent's capacity before attempting to measure the values of SC. 

\subsubsection{Relations Measure} \label{sec:relation}

In order for agents to tackle different problems for the continuation of their organization, they form a subgroup that best fits for a given problem. Even though their relations have a huge impact on the formation as well as the coordination in this world, subgroup formation as well as task or problem allocation is outside the scope of this work. We focus on measuring a network of relationships for subsequent determinations of different values that an agent accumulates when interacting with others. The initial values of relations are provided by every agent when she first joins an organization. Those relations and their values are not static and agents are able to create, diminish, or improve each one of them depending on current actions and interaction.


For every action an organization performs, there exists a goal $G_j \in \{G\}$. Each goal will be distributed into a set of tasks, such that $G_j \to \{\Theta\} = \langle \theta_1, \dots, \theta_n \rangle$, for possible assignments to agents. The completion of one task $\theta_m \in \{\Theta\}$ includes interaction between agents for a set of subtasks $\{\theta_m\}= \langle \theta_m^1, \dots, \theta_m^k \rangle$. The coordination as well as control of those tasks are determined by the organization. We benefit from the dynamic interaction among agents while achieving multiple tasks in order to update the current values of relations. Those values of relations depends on the nature of interaction over every given task; therefore, we model relations in a task-based scenario to describe the continual changes over time in inter-agent connections and to help with updating relations throughout repeated task assignments. In the case of OSS Project, the goal is to develop the project and the sub-goals are the releases of the project. The tasks are the software modules that need to be developed in order to achieve a sub-goal (i.e., releasing the project). The subtasks of one module are the lines of codes to be added or deleted in order to complete the module. The interaction between agents (contributors) working in a task (module) occur through the completion of the subtasks.


For every self-selected task, agents define a task-based graph upon the initial relations and there is at least one active
edge that prescribes a plan. Agents are able to form an edge through successive interaction. 
In other words, the network structure allows for the property of transitivity 
which permits interaction over that edge to improve giving it the chance to reach a threshold in order to be
considered active. Interaction are commonly observed of two types of affinities~\cite{Smith:2009}, where 
(a) \textit{explicit affinities} become evident through interaction over an existing relation (i.e. it is observed when two or more agents have interaction with whom they have a previous experience over an existing edge in the graph), and 
(b) \textit{implicit affinities} allow for other possible interaction among agents without previously modeled relations. In the case of OSS Project, explicit affinities between two agents exist when both agents contributed on a common software modules. 
Interaction emerge from the closure property of relations~\cite{Granovetter:1973} and may help in forming new edges when updating relations (i.e. previously un-modeled relations).


The current values of relations are updated every time interval $\Delta t$. 
For the general assembly, we describe existing relations as explicit links; otherwise, they will be considered as implicit. Values of links are proportional to the frequency of interaction over them. The value on an explicit edge, $\mathcal{L}$, between agent $i,i' \in N$, is computed accumulatively based on the frequency of interaction between the two agents throughout the time interval (i.e. $\Delta t = t_2 - t_1$). 
%
%
This is stated in Equation~\ref{eq:exl} at a specific subtask $\theta_m^s$, where $t_r$ is the end of duration that spans from $t_1$ toward $t_2$. 
\begin{equation} \label{eq:exl}
\mathcal{L}_i^{i'} (\theta_{m}^s, t_2) = \mathcal{L}_i^{i'} (\theta_{m}^s, t_1) + \sum_{r \in \Delta t} I_{nteractions}^{i,i'} (\theta_{m}^s, t_r)
\end{equation}
\hfill where $ \theta_m^s$ is a subtask $\in \{\theta_m\}$, $\forall  i \neq i' \in N$, and $t$ for time.
%

Implicit links are traditionally observed through triadic closure theory \cite{Granovetter}. 
Triadic closure, in short, asserts that for each three agents $i$, $i'$ and $i''$ where two explicit affinities exist in term that link $i \leftrightarrow i'$ and $i' \leftrightarrow i''$ , there should exist an implicit affinity that links $i \leftrightarrow i''$. In a triadic formation of two explicit affinities, there are different possibilities for the value $\in \mathbb{R}$ that the implicit affinity should have. 
The possible value that an implicit affinity may obtain depends on the value of the current explicit edges.
%
Thus, we can state that the initial value of the third implicit link $\mathcal{L}_i^{i''}$ in a triad can be approximated in Equation~\ref{eq:iml}. 

\begin{equation} \label{eq:iml}
\mathcal{L'\:}_i^{i''} (\theta_m^s) \equiv \frac {\mathcal{L}_i^{i'} (\theta_m^s)+ \mathcal{L}_{i'}^{i''}(\theta_m^s)} {\big| R_{elation}^{i,{i'}} (\theta_m)+R_{elation}^{{i'},{i''}} (\theta_m)\big|^2} 
\end{equation}
\hfill where $ \theta_m^s$ is a subtask $\in \{\theta_m\}$ and $\forall  i \neq i' \neq i'' \in N$

We are considering the formation of implicit links through explicit links only. That means there must be an explicit path from the source node to target node in order for an implicit link to exist. The traversal in the path of unrepeated explicit links between $i$ and $i'$ will consider the maximum volume despite distances. An extension of the closure envisioned in (\ref{eq:iml}), where there existed two disjoint (i.e. nonconsecutive) links with explicit affinities or possible undefined links in between, is determined through Equation~\ref{eq:iml2}.
\begin{equation} \label{eq:iml2}
\mathcal{L'\:}_i^{i'} (\theta_m^s) \equiv \frac{1}{\big| \sum_{i,i'} R_{elation}^{i,i'} (\theta_m) \big|^2}  \cdot \sum_{i,i'} \mathcal{L}_i^{i'} (\theta_m^s)
\end{equation}
\hfill where $ \theta_m^s$ is a subtask $\in \{\theta_m\}$, and $\forall  i \neq i' \in N$.

Agents' interaction are instrumental in forming new implicit links and updating the values of existing explicit ones. During a task completion, it is possible for frequently used implicit relations to gain a sense of actualization; thereby, the implicit relations will be treated the same as explicit ones. Next, we model relations considering those measurements of explicit as well as implicit links. As stated earlier, the initial values of relations are provided by the agents' public profiles and are used in forming a task-based socio-graph. We mapped those relations to explicit links in a task-based graph in order to capture current interaction as well as to allow possible measures of implicit links. By the time a new task is going to be assigned, an organization updates agents' relations over all tasks based on the new values of links. When a relation from an implicit link ($\mathcal{L'}$) reaches a threshold value of $\tau$ that has been specified previously by an organization, it will be treated as an explicit one and an agent is able to explicitly form a relation over it. It is possible for those relations to have a value of positive, negative, or mutual (equal) for non-existing or possible unprejudiced relations. Agents will update their profiles as well using those new relations values for a later possible assignment.

Equation~\ref{eq:relation2} updates the initial value of relation between every pair of agents by considering the most repeated value over an explicit or an implicit link at a given subtask. 

\begin{equation} \label{eq:relation2}
R_{elation}^{i \to i'} (\theta_m) = \mode_s \; \big( \mathcal{L}_i^{i'} (\theta_m^s) + \mathcal{L'}_i^{i'} (\theta_m^s) \big)
\end{equation}
\hfill where $\forall \; i \neq i' \in N$ and $\forall \; \theta_m^s \in \{\Theta\}$.

\vspace{-2mm}
\subsubsection{Capacity Measure} \label{sec:capacity} 
Agent's capacity can be described as the absolute ability to accomplish tasks given the time constrains and interests. A measurement of an agent's capacity is a critical issue and should be addressed once an agent joins an organization. This will eliminate the possibility of agent's ineligibility to accomplish tasks when allocated to it. 
The value of capacity is dynamic and rapidly changing from one task to another. For simplicity, we consider capacity to be a combination of an agent's innate 
(1) $\mathsf{capabilities}$ for the ability to achieve different tasks, extemporaneous 
(2) $\mathsf{willingness}$ to perform certain actions based on her preferences, and ad-lib
(3) $\mathsf{availability}$ for her readiness to participate.
Agents' capabilities and willingnesses are provided in their public profiles while availabilities are ranging from $[0\to1]$ based on the task they occupy. 
Willingness is the degree of commitment to which an agents is ready to work hard to achieve the organizational objectives. 
The willingness of an agent is important in determining her contributions for a task. 
Equation~\ref{eq:capacity} shows a very direct measurement for agent $i$'s capacity to achieve a certain task $\theta_m$. 
\begin{equation}\label{eq:capacity}
\begin{split}
C_{apacity}^i (\theta_m) = \Big(\mathsf{capability}_i (\theta_m) + \mathsf{willingness}_i (\theta_m)\Big)  
\times \; \mathsf{availability}_i (\theta_m)
\end{split}
\end{equation}
\hfill where $\theta_m \in \{\Theta\}$ and $\forall \; i \in N$

Due to the rapid changes in the agent's capacity, an agent will not be able to preserve them for future use. They must be updated instantaneously every time a new task is assigned. We assume that the capacity of an agent is independent $\forall i \in N$.

\vspace{-1mm}
\subsubsection{The value of benevolence}

Agents entering an organization and interacting with those whom they have no previous interaction are initializing their benevolent values with a constant of a \textit{Null}; then, the benevolences are derived from their relationships with others as well as their capacities to overcome certain problems. 
%
%
Due to the fact that an organization is a formation that overlay a dynamic network, we model benevolence between agents based on a directed network's graph of connected vertices and edges. The resultant graph will be a task-based weighted graph of vertices as agents capacities and edges as their relations. 
The weighted benevolent graph is connected, and there should at least be one active relation between any pair of agents. We follow next with a formal definition of the weighted benevolent graph while emphasizing on the parameters that contribute to its value. 

Let $N \subseteq \mathcal{N}$ be set of agents working on a goal $G_i$. There exists a set of tasks (i.e. 
$\{\Theta\} = \langle \theta_1, \;\dots,\; \theta_m \rangle$) for each goal. 
Let $w:2^N \to x$, where $w(N) \in \mathcal{N}$ is a world of $N$-agents working on $\theta_m$, and $x$ is a random variable with distribution that has not been determined yet. The parameters of the $w(N)$ are attained from an organization and sampled over existing $k$-subtasks to all $N$. 
Let $B_{enevolence}^i : \mathbb{R}^k \to \mathbb{R}$ be the benevolent function of real values that computes the benevolent value of $w(N)$ at $\theta_m$ based on the distribution of $k$-sub-tasks. 
We are trying to find out the benevolent values resulting from unilateral relationships between agents of $N \subseteq \mathcal{N}$ in the $w(N)$.


A benevolent socio-graph is basically a combination of agents and relations. 
The value of relations can be different from one task to another; however, for the sake of simplicity, we will be evaluating those relations in a task-based graph. 
We use the normal distribution to correspond to the average values of agents benevoelnces with a peak and the variability with other agents in a symmetric spread. 
i.e. $C_{apacity}^i=(C_{apacity}^{i,1},\; \dots ,\; C_{apacity}^{i,k})$, where $k$ is number of subtasks and $C_{apacity}$ is the agent's capacity $\forall \; C_{apacity}^{i,k}  \sim C_{apacity}(\mu_{i,k}  ,\sigma_{i,k}^2)$. The benevolence between a pair of agents $(i, i')$ can be presented in Equation~\ref{Eq:ben}.
\begin{equation} \label{Eq:ben}
B_{enevolence}^{i \to \neg i} (\theta_m)= R_{elation}^{i \to \neg i} (\theta_m) \cdot C_{apacity}^i (\theta_m)
\end{equation}

The values of relations are critical in this case, they are resulting from a weighted directed graph of the network. The benevolence takes advantage of agents' current relations and the rapid changes in their values within the assignment of one task. We take into consideration an agent current interests and readiness to contribute captured in the measurement of capacity. 
Although implicit links are not considered when defining benevolence, current values of relations have already considered them, and they will directly contribute to current values of benevolence once a specific threshold is reached. 

\vspace{-1mm}
\subsubsection{The value of potential-benevolence}
Agents' beliefs play an important role in the expected receipt of SC. When an agent {believes} that another is able to provide resources to her, she will then try to obtain those resources. When resources are obtained, trust is initiated. Agents providing resources are then of higher power and importance than the agent acquiring them. Since the value of the SC that initialize the link from acquirer to provider is proportional to the acquirer belief, we consider belief to be a function of the directed link to the provider. The value obtained from this function is proportional to the value of the SC gained by the provider. Given a graph of $N$-nodes and $i$ is one of the nodes while $\neg i$ are other member nodes $\in N$ that are $\neq i$, the potential benevolence of agent $i$ receiving a contribution from other agents within $N$ is obtained through Equation~\ref{SC:PB}. 

\begin{equation} 
\begin{split}
PB_{enevolence}^{i} (\theta_m) = \sum_{\forall \neg i \in N} B_{elief}^{\neg i} \big( C_{apacity}^i (\theta_m)\times R_{elation}^{\neg i \to i} (\theta_m) \big)
\label{SC:PB}
\end{split}
\end{equation}   

Equation~\ref{SC:PB} states that the value of an agent's capacity is a critical parameter for receiving a benevolence. The value of a relation from $i \to \neg i$ is not the sum of all links an agent traverses through to get to the provider. It can be calculated through an implicit link $\mathcal{L}'$ if an explicit directed link, i.e. $\mathcal{L}$, is not available. 

\subsection{A measurement of social capital}
The SC for an agent is based on her beliefs to receiving contribution from peers over the network. 
%
%
The probability of an agent providing a continual benevolence to another is proportional to the expected capacity that the acquirer may be interested in, as stated in Equation~\ref{SC:believe-to-link}. We are considering, in our measurements of SC, a task-based graph for that the following equations are for a specific-task (e.g. $\theta_m$).
\begin{equation} 
%
\resizebox{0.91\hsize}{!}{
$f \big( B_{enevolence}^{\neg i \to i} | PB_{enevolence}^{\neg i \to i} \big) = 
  \begin{cases}
    \sum_{\forall \neg i \in N}   \Bigg[ \frac{B_{enevolence}^{\neg i \to i} \cap PB_{enevolence}^{\neg i \to i}}{PB_{enevolence}^{\neg i \to i}} \Bigg]      & \quad \text{if } PB_{enevolence} > \text{ 0}\\
    0  & \quad \text{ Otherwise;}
  \end{cases}
  $}
\label{SC:believe-to-link}
\end{equation}   

Intersection means the expectation to receive benevolence considering the given benevolence; otherwise zero. In our case, we can eliminate the value of potential benevolence after the intersection and assume that the value of benevolence is true if the potential one exists. Equation~\ref{SC:value} shows that directed SC is gained by the provider agent.

\begin{equation} \label{SC:value}
SC_{i} = \sum_{\forall \neg i \in N} B_{elieve}^{i} \bigg( \; f \big( B_{enevolence}^{\neg i \to i} | PB_{enevolence}^{\neg i \to i} \big) \; \bigg)
\end{equation}   

Belief is a {decay} function that decreases the value of SC received when traversing {through} multiple agents. It is exponential to how many explicit links the acquirer has to travel through to obtain resources from the agent provider. 
We introduce the belief function $B_{elief}: \mathbb{R}^+ \to \mathbb{R}^+$, where $B_{elieve}^{i \to \neg i}(R_{elation}^{i \to \neg i})$ is the belief of the relation that returns the task based between agents $i$ and $\neg i$. Belief is a monotonically decreasing 
function so that a larger number of relations corresponds to a lower belief. The belief value is domain specific and an example of it can be: 
%
$B_{elieve}^i = e^{- \lambda \cdot ( R_{elation}^i )}$.

When an agent capitalizes on another, her current capacity is also accessible for that agent to take advantage form - in-return. When both agents capitalize on each others, they form a cooperative behavior that contributes positively to the organizational SC feeding back to the organization member-agents.







\section{The case of an OSS project}
Open source software (OSS) is a type of software projects whose source code is released. The users of an OSS system may have the right to change source code of system. The development process of OSS projects are different than industrial software projects. OSS development are based on collaborations between multiple independent developers (aka contributors) who aim to achieve a common goal. The contributors are usually staying in different geographical areas. Thus, OSS projects mostly have online repositories that allow multiple developers to contribute independently to the project~\cite{mendez2009returns}. Over the last two decades, OSS development have gained popularity and we have witnessed successful OSS projects such as Linux, MySQL, and Hadoop. However, the majority of OSS projects have failed due to different reasons. 
In this paper, we try to understand the impact of SC on OSS development and whether it has a relation with the success of OSS projects.


We consider the OSS Project \texttt{Hadoop} as a case study to illustrate the SC value computation. The data are taken from GitHub portal from year 2013 to 2018, and we divide the data into three time intervals (2013-2014, 2015-2016, and 2017-2018). In this case study, the task is a package in the project and the sub-task $\theta_m^s$ is a class in the package. We select an \texttt{Apache-Hadoop} package\footnote{\texttt{org.apache.hadoop.yarn.client.api}} as an example that has $10$ classes. There are $20$ contributors involved in the package development. We consider the number of line code as the value of interaction among the contributors (adding lines and deleting lines). Table~\ref{table:1} shows an example of data collected from the GitHub portal on a class named \texttt{AMRMClient.java}. $15$ contributors involve in the class development. 
%
%
For simplicity, the capacity of a contributor is defined as the total number of commits the contributor conducted during the considered three time intervals and the value of the interaction of the contributor with the others (i.e., the other contributors who work with the contributor in same class) is the sum up of number of lines of code that the contributor added and deleted in the class.

\vspace{-7mm}
\begin{table}[htb]
\caption{Measurements of SC for a subgroup of three developers in an OSS project}
\label{table:1}
\centering
\resizebox{\columnwidth}{!}{
\begin{tabular}{c|c|ccc|cc|cc|c}
 \hline
	 \multicolumn{10}{|c|}{$G_1$}	\\
 \hline
Interval 	& 	Agents	& 	Links 	& 	Relation &		Capacity	&  Benevolence 	&	PBenevolence	& instant-SC		&	Accumulative-SC	&  Net-SC	\\
 \hline
		&	Vin	&	204	&	204	&	21	&	4284	&	980		&	4.371 	&	4.371	&\\
$t_1$	&	Oz	&	190	&	190	&	5	&	950	&	4314		&	0.221 	&	0.221	&4.598\\
		&	Roh	&	10	&	10	&	3	&	30	&	5234		&	0.006 	&	0.006	&\\
		\hline
		&	Vin	&	367	&	367	&	21	&	7707	&	1831		&	4.209	&	8.58		&\\
$t_2$	&	Oz	&	365	&	365	&	5	&	1825	&	7713		&	0.236	&	0.457	&9.044\\
		&	Roh	&	2	&	2	&	3	&	6	&	9532	 	&	0.001	&	0.007	&\\
		\hline
		&	Vin	&	301	&	444	&	21	&	9324	&	2185		&	4.267	&	12.847	&\\
$t_3$	&	Oz	&	103	&	401	&	5	&	2005	&	9504		&	0.211	&	0.668	&13.537\\
		&	Roh	&	60	&	60	&	3	&	180	&	11329	&	0.015	&	0.022	&\\
 \hline
\end{tabular}
}
\end{table}
\vspace{-6mm}

Table 1 shows the empirical results on calculating the social capital of three developers involved in the OSS projects. The SC values change accordingly after the group of developers finished one task after another. The value of SC is shown to have a positive correlation to the agent productivity and net capital for releasing the software. This exemplification and verifications of the proposed measurements for SC, in part, supports the hypothesis that is an increase in SC leads to a better teamwork for an optimized productivity that is observed here through new/advanced software releases.

 \section{Conclusions and Future work}\label{Conclusion} \vspace{-4mm}
In open environments, agents are integrated dynamically across their organizational and geographical boundaries to justify each other's needs. Such systems should be modeled as an open multi-agent system, in which semi-autonomous agents can interact in an open environment, despite potentially conflicting interests. 
Agents collaboration for a long term objectives leads to forming an organization that is commonly subservient to electronic institutions. Water quality and health quality exchanges are two examples of electronic institutions guiding large subordinate organizations. Electronic institutions must routinely monitor and improve SC by updating the organization's policies. %
Electronic organization is part of electronic institutions, we can have a policy about SC of electronic institutions among electronic organizations (e.g. World Bank and UN). %

This paper defined SC 
and proposed a measurement for it based on the benevolences between autonomous agents operating in a large-scale open service-oriented organization. Incorporating benevolence, in measuring the social capital for individual agent and for the organization as whole, gives more tangible values. Those values contribute positively towards the cooperative nature of an organization. We showed an empirical evaluation of the proposed approach using a real-world case study of an open-source project development. 
The future work will propose a detailed SC assessment model that is required to estimate the future behavior of agents and agents' peers in order to simplify the interaction process with those peers. 

 \bibliographystyle{splncs04}
\bibliography{citations-complete}

\begin{thebibliography}{10}
\providecommand{\url}[1]{\texttt{#1}}
\providecommand{\urlprefix}{URL }
\providecommand{\doi}[1]{https://doi.org/#1}

\bibitem{Adler:2000}
Adler, P.S., Kwon, S.: Social capital: The good, the bad, and the ugly. In:
  Lesser, E.L. (ed.) Knowledge and Social Capital, pp. 89 -- 115.
  Butterworth-Heinemann, Boston (2000)

\bibitem{Bonacich:87}
Bonacich, P.: Power and centrality: A family of measures. American Journal of
  Sociology  \textbf{92}(5),  1170--1182 (1987)

\bibitem{Bourdieu:1986}
Bourdieu, P.: The forms of capital. In: J.~Richardson, t.R.N. (ed.) Handbook of
  Theory and Research for the Sociology of Education. pp. 241--258. Greenwood
  Press, New York (1986)

\bibitem{Burt:2000-NetSC}
Burt, R.S.: The network structure of social capital. Research in Organizational
  Behaviour  \textbf{22},  345 -- 423 (2000)

\bibitem{burt2005brokerage}
Burt, R.S.: Brokerage and closure (2005)

\bibitem{Granovetter}
Granovetter, M.: The strength of weak ties: A network theory revisited, pp. 105
  -- 130. Sage, Beverly Hills, CA (1982 1982)

\bibitem{Granovetter:1973}
Granovetter, M.S.: The strength of weak ties. American Journal of Sociology
  \textbf{78}(6),  1360 -- 1380 (1973)

\bibitem{Gummer:1998}
Gummer, B.: Social relations in an organizational context: Social capital, real
  work and structural holes. Administration in social work  \textbf{22}(3),  87
  -- 105 (1998)

\bibitem{hsu}
Hsu, J., Hung, Y.: Exploring the interaction effects of social capital. Inf.
  Manage.  \textbf{50}(7),  415--430 (2013)

\bibitem{lin1999building}
Lin, N.: Building a network theory of social capital. Connections
  \textbf{22}(1),  28--51 (1999)

\bibitem{lin2002social}
Lin, N.: Social capital: A theory of social structure and action, vol.~19.
  Cambridge university press (2002)

\bibitem{mcpherson2001birds}
McPherson, M., Smith-Lovin, L., Cook, J.M.: Birds of a feather: Homophily in
  social networks. Annual review of sociology  \textbf{27}(1),  415--444 (2001)

\bibitem{mendez2009returns}
M{\'e}ndez-Dur{\'o}n, R., Garc{\'\i}a, C.E.: Returns from social capital in
  open source software networks. Journal of Evolutionary Economics
  \textbf{19}(2),  277--295 (2009)

\bibitem{Nahapiet:1998}
Nahapiet, J., Ghoshal, S.: Social capital, intellectual capital, and the
  organizational advantage. The Academy of Management Review  \textbf{23}(2),
  242--266 (1998)

\bibitem{ostrom:2003}
Ostrom, E., Ahn, T.: Foundations of Social Capital. Critical studies in
  economic institutions, Edward Elgar Pub. (2003)

\bibitem{Pitt2016}
Pitt, J.: From Trust and Forgiveness to Social Capital and Justice: Formal
  Models of Social Processes in Open Distributed Systems, pp. 185--208.
  Autonomic Systems, Springer International Publishing (2016)

\bibitem{Smith:2009}
Smith, M., Giraud-Carrier, C., Purser, N.: Implicit affinity networks and
  social capital. Inf. Technol. and Management  \textbf{10}(2-3),  123 -- 134
  (2009)

\end{thebibliography}

\end{document}